# Two-dimensional gap solitons in elliptic-lattice potentials

Yingji He,*[1,2] Boris A. Malomed,[3] and Bambi Hu[1,4]

[1]*Department of Physics, Centre for Nonlinear Studies, and The Beijing-Hong Kong-Singapore Joint Centre for Nonlinear and Complex Systems, Hong Kong Baptist University, Kowloon Tong, Hong Kong, China*

[2]*School of Electronics and Information, Guangdong Polytechnic Normal University, 510665 Guangzhou, China*

[3]*Department of Physical Electronics, School of Electrical Engineering, Faculty of Engineering, Tel Aviv University, Tel Aviv 69978, Israel*

[4]*Department of Physics, University of Houston, Houston, Texas 77204-5005, USA*

*Corresponding author, E-mail address: yjhe@hkbu.edu.hk

**Abstract**

We study two-dimensional (2D) matter-wave gap solitons trapped in an elliptically deformed concentric lattice potential, within the framework of the Gross-Pitaevskii equation (GPE) with self-attraction or self-repulsion. For a fixed eccentricity of the lattice, soliton families are found in both the repulsive and attractive models. In the former case, the analysis reveals two kinds of gap solitons trapped in the first oval trough (the ring-shaped potential minimum closest to the center): *elliptic annular solitons* (EASs), and *double solitons* (DSs), which are formed by two tightly localized density peaks located at diametrically opposite points of the trough, with zero phase difference between them. With the decrease of the norm, the density distribution in the EAS along the azimuthal direction changes from nearly-uniform to double-peaked and, eventually, to the DS. In the attractive model, there exist only DSs in the oval trough, while EASs are not found. All such solitons without the angular momentum ($l = 0$) are fully stable. For $l \neq 0$, *vortical solitons* – both EASs with a sufficiently large norm (in the repulsive model) and DSs (in models with both signs of the nonlinearity) – are quasi-stable, exhibiting rocking motion in the elliptic trough (we consider the cases of $l=1$ and $l=2$). At smaller values of the norm, the vortical annular solitons (in the repulsive model) are unstable. Stable fundamental solitons trapped in the central potential well are investigated too, in both the attractive and repulsive models, by means of the variational approximation and numerical methods.

PACS number(s): 42.65.Tg, 42.65.Jx, 42.65.Wi

Keywords: spatial solitons, elliptic lattices, Bose-Einstein condensates, matter-wave solitons

## I. INTRODUCTION

Techniques which make it possible to manipulate matter-wave beams by means of specially designed trapping and stirring potentials are important ingredients in many applications of atom optics, such as inertial sensors [1] and atomic holography [2]. One of essential topics appearing in that context is the control of the dynamics of Bose-Einstein condensates (BECs) trapped in toroidal magnetic [3] and optical waveguides [4].

Optical lattices (OLs), i.e., periodic potentials induced by laser beams illuminating the experimental field, is a versatile tool in the studies of ultracold quantum gases [5-11]. In particular, matter-wave solitons of the gap type, i.e., those whose chemical potential falls into a bandgap of the linear spectrum induced by the OL, were predicted [5] and created [6] in BECs with repulsive interactions between atoms. An especially promising application of the OLs is the stabilization of multidimensional solitons in BEC [7]. In the latter context, it was also predicted that two-dimensional (2D) solitons and solitary vortices can be stabilized by a revolving rectangular or quasi-1D OL, for either sign if the intrinsic nonlinearity, attractive or repulsive [8]. Previously, it had been demonstrated, in the case of the self-attraction, that axisymmetric lattice potentials of the Bessel type support stable tightly localized 2D *rotary solitons*, which may perform circular motion in an annular potential trough [9]. The same model supports stable 3D solitons [10]. If the nonlinearity is repulsive, the Bessel radial potential maintains 2D ring solitons in the form of azimuthal dipoles and quadrupoles [11]. Similar settings are possible in nonlinear optics. In particular, spatial optical solitons localized at the center or forming a ring, as well as rotating ones, were created in photoinduced concentric lattices in a photorefractive material [12].

It is pertinent to mention that various complex (delocalized) multidimensional patterns based on vortices were also theoretically investigated in BEC models based on the Gross-Pitaevskii equations (GPEs) with the repulsive nonlinearity [13]. Another ramification of these studies has demonstrated that an axisymmetric lattice potential, periodic along radial variable $r$ (rather than decaying at $r \to \infty$, as in the case of the above-mentioned Bessel lattice), gives rise to stable 2D localized modes, in the form of both fundamental gap solitons trapped at the center of the potential, and vortex solitons trapped in lattices rings [14]. A somewhat similar model was elaborated in nonlinear optics, for annular gap solitons in the case of the radial light transmission in a disk-shaped nonlinear waveguide equipped with the circular Bragg grating [15].

In addition to circular solitons and vortices, their non-axisymmetric (elliptically shaped) counterparts were theoretically predicted and for the first time experimentally observed in a photorefractive medium with an anisotropic nonlocality [16], and later created in the local photorefractive material [17]. Modes of this kind were also investigated in BECs [18].

Recently, it has been demonstrated that elliptically shaped vortex solitons in a defocusing nonlinear medium, with composite "Mathieu lattices" imprinted into it, feature patterns which are anisotropic both in the intensity and phase, provided that their integral power exceeds a certain threshold value [19].

In this work, we study 2D solitons supported by elliptic counterparts of the above-mentioned circular lattice periodic along $r$, in cases of both the self-attraction and self-repulsion. In terms of the 2D polar coordinates, $(r,\theta)$, the respective elliptic-lattice potential with strength $p$ may be written as

$$V(r) = -p\cos\left(2r\sqrt{1+\delta\cos(2\theta)}\right), \tag{1}$$

where parameter $\delta$, with $0 \leq \delta < 1$, determines the ellipticity (eccentricity) of the potential. Examples of elliptic lattices are displayed in Fig. 1. This setting can be readily created in the experiment, by illuminating a "pancake"-shaped BEC layer by an oblique radially –structured optical beam (the same as one which creates the axisymmetric radial OL [14], in the case of the normal incidence). In optical media with an imprinted material radial grating, its elliptic variety may be realized by an anisotropic deformation of the sample.

Here, we demonstrate that the elliptic-lattice potential gives rise to new dynamical effects,

such as rocking motion of vortical annular solitons. This setting also makes it possible to consistently introduce the concept of the vorticity in the elliptic geometry, even though the latter does not provide for the conservation of the angular moment.

In the repulsive model, we demonstrate the existence of a threshold value of the norm for a fixed ellipticity, above and below which elliptic annular solitons (EASs) or double solitons (DSs), are formed, respectively, in the first elliptic (oval) potential trough. The DSs are built as pairs of in-phase density maxima located at diametrically opposite points. In EASs, the density distribution is almost uniform along the trough if the norm is large enough, the distribution assuming a saddle shape with the decrease of the norm. In the attractive model, only DSs exist in the oval troughs. The fundamental solitons, i.e., those with zero vorticity $l$, are fully stable. At $l=1$ and $l=2$, both the vortical EASs with a sufficiently high norm, in the model with the repulsive nonlinearity, and vortical DSs, with the nonlinearity of either sign, are quasi-stable. Stable fundamental solitons trapped in the central potential well are analyzed in this work too.

The paper is organized as follows. The model is formulated in Section II. Solitons of the EAS and DS types, trapped in the first annular trough of the elliptic lattice are considered in Section III – in detail for the repulsive model, and in a brief form in the model with the self-attraction. Section IV is dealing with solitons trapped in the center of the elliptic lattice, for both signs of the nonlinearity. The paper is concluded by Section V.

## II. THE MODEL

We use the GPE for the 2D BEC for wave function $u$ in the usual scaled form [14,19],

$$i\frac{\partial u}{\partial t} = -\left(\frac{\partial^2}{\partial r^2} + \frac{1}{r}\frac{\partial}{\partial r} + \frac{1}{r^2}\frac{\partial^2}{\partial \theta^2}\right)u + V(x)u + \alpha |u|^2 u = 0, \qquad (2)$$

where $\alpha = +1$ and $\alpha = -1$ correspond to the repulsive and attractive nonlinearity, respectively, and the potential is taken as specified in Eq. (1). This elliptic-lattice potential is uniform along elliptic rings, $r^2[1+\delta\cos(2\theta)] = \text{const}$, which makes it different from the Mathieu-lattice potential [19]. The latter one is non-uniform, featuring minimum and maximum at the minor and major axes, respectively. The diffraction of optical beams inducing potential (1) in the case of the 2D BEC model is not essential, as the respective diffraction length is much larger than the thickness of the BEC layer in the possible experiment [14]. In the application to nonlinear optics, Eq. (2) is the nonlinear Schrödinger equation for the propagation of light in a nonlinear bulk medium equipped with the elliptic grating, variable *t* being the propagation distance.

Soliton solutions to Eq. (2) are looked for in the form of

$$u(r,\theta,t) = \phi(r,\theta)e^{-i\mu t + il\theta}, \qquad (3)$$

where real $\mu$ is the chemical potential, integer *l* is the vorticity (we consider values $0 \leq l \leq 2$), and radial function $\phi(r,\theta)$ obeys equation

$$\frac{\partial^2 \phi}{\partial r^2} + \frac{1}{r}\frac{\partial \phi}{\partial r} + \frac{1}{r^2}\left(\frac{\partial}{\partial \theta} + il\right)^2 \phi + \left[\mu + p\cos\left(2r\sqrt{1+\delta\cos(2\theta)}\right)\right]\phi - \alpha|\phi|^2\phi = 0. \qquad (4)$$

The norm of the solution, which is proportional to the total number of atoms in the BEC, or total power carried by the beam in the optical model, is $N = \int_0^{2\pi} d\theta \int_0^\infty r dr |\phi(r,\theta)|^2$. Solutions for solitons and solitary vortices were obtained by means of the well-known method based on beam-propagation simulations of Eq. (2) in imaginary time.

## III. ANNULAR AND DOUBLE ELLIPTIC SOLITONS

**A. The repulsive model**

In the repulsive model, with $\alpha = +1$, the elliptic lattice can support fundamental and vortical EASs and DSs in the first ring-shaped trough, as shown in Fig. 2. They may be regarded as counterparts of the radial gap solitons which were studied in the circular lattice potential [14]. The gap character of such solitons is established by verifying (not shown here in detail) that the corresponding values of chemical potential $\mu$ [see Eq. (3)] belong to the first finite bandgap of the 1D linear spectrum induced by the quasi-flat OL potential far from the central region, cf. Ref. [14].

Typical examples of the solitons are displayed in Fig. 3. Note that the local density is nearly uniform along the oval trough in the EASs whose norm is large enough [see Figs. 3(a) and 3(d)], while vortex solitons supported by the elliptic Mathieu-lattice potential feature an anisotropic annular distribution, irrespective the norm [19]. The same is true concerning the local phase gradient (superfluid velocity), which is nearly uniform along the annulus in the present model (not shown here in detail), and strongly anisotropic in the case of the Mathieu lattice. The latter model features the largest and smallest tangential phase gradients near the minor and major elliptic axes, respectively, which implies a possibility of the creation of vortices via a bifurcation from dipole modes. This does not occur in the present model: as seen in Fig. 2, the type of the soliton is mainly determined by the interplay of norm *N* and ellipticity *δ* of the lattice potential.

With the decrease of the norm, the EASs exhibit a transition to DSs via saddle-shaped distributions of the density along the azimuthal direction [see Figs. 3(b) and 3(e)], with maxima and minima collocated, respectively, with major and minor axes of the elliptic potential. Indeed, it is obvious that segments of the oval trough close to the major axis, which feature the largest curvature, may locally trap a larger norm. Continuing this trend, a distinct feature of the DS is that the local density vanishes at points where the minor elliptic axis intersects the oval trough, while two density peaks coincide with the position of the major elliptic axis. With the increase of the ellipticity at a fixed norm, the EAS also transforms into the DS. The critical values of the ellipticity and norm, that may be identified as corresponding to the transformation of the soliton from EAS to DS, are indicated by the boundary between areas A and B in Fig. 2. As shown in Figs. 4, 5, and 6, the existence and stability regions for the EASs and DSs with vorticities $l = 0,1$, and $2$, respectively, were also identified in the plane of $(p,\mu)$.

The stability of the solitons was tested by way of direct simulations of Eq. (2) with initial conditions perturbed by a random noise at the level of 10% of the soliton's amplitude (actually, this is strong noise). The simulations demonstrate that the fundamental EASs with $l = 0$ are stable in the entire region of their existence, while for $l \neq 0$, EASs need the norm large enough to maintain their effective stability, see examples in Figs. 4(b), 4(c), 5(b), 5(c), 6(b) and 6(c). The minimum norm necessary for the stability of the vortical EASs increases with the ellipticity, see regions B and C in Fig. 2. The stabilization of the annular solitons with the increase of $N$, i.e., the effective strength of the self-repulsion, is quite natural [14,15]. Actually, the character of the stability of the elliptic annular gap solitons is similar to that reported for the circular lattice in Ref. [14]. In particular, the vortical EASs with $l = 1$ and $l = 2$, whose norm is large enough, do not maintain an exactly stationary shape. Instead, they exhibit rocking motion in the elliptic ring, periodically switching between the anti-clockwise and clockwise directions, similar to the motion of the DSs presented in Figs. 5(e) and 6(e). In the circular geometry, slightly deformed annular vortex solitons rotate in one direction [14], the rocking in the present case being caused by bounces from "narrow necks" in the oval trough. In regions B and C of Fig. 2, unstable EASs with $l = 1$ and $l = 2$ suffer strong losses due to the rotational motion, and eventually split into fragments (two fragments in the case of $l=1$, and several of them, if $l=2$).

As said above, the diagram shown in Fig. 2 indicates a transition to DSs with the further decrease of the norm. The DSs trapped in the oval trough are stable for $l=0$, and exhibit a quasi-stability, in the form of the above-mentioned rocking motion, for $l \neq 0$ [see Figs. 5(e) and 6(e)]. We stress that the density maxima in the DS have zero phase difference, in contrast with formally similar dipolar modes trapped in the Mathieu lattices [19], in which the phase difference between the maxima is $\pi$.

**B. The attractive model**

In the model with the self-attraction ($\alpha=-1$), the oval trough supports DSs only, while solutions of the EAS type could not be found, irrespective of the norm. There is a maximum critical value of the norm, beyond which solitons do not exist due to the possibility of the collapse, see Fig. 7. Two examples of DSs are shown in the Fig. 8 for different depths of the lattice. In particular, stronger lattices support the DSs with narrower and taller profiles, see Fig. 8(b). These DSs turn out to be stable and quasi-stable for $l=0$ and $l \neq 0$, respectively. In the latter case, the DSs exhibit the rocking motion, similar to that in the repulsive model – cf. Figs. 5(e) and 6(e). In terms of the 1D linear spectrum induced by the quasi-flat OL potential far from the center, all the solitons found in the attractive model belong to the semi-infinite gap.

**IV. FUNDAMENTAL SOLITONS TRAPPED AT THE CENTER OF THE ELLIPTIC LATTICE**

Next, we use the variational approximation (VA) to analyze the existence of localized modes trapped in the central potential well (around $r=0$), in the repulsive and attractive models alike. To this end, we introduce an *elliptical ansatz*,

$$u(r,t) = A(t)\exp\left\{\frac{-\left[r\sqrt{1+\delta\cos(2\theta)}\right]^2}{2[a(t)]^2} + \frac{i}{2}b(t)\left[r\sqrt{1+\delta\cos(2\theta)}\right]^2 + i\phi(t)\right\}, \quad (5)$$

with amplitude $A$, width $a$, chirp $b$, and overall phase $\phi$. Inserting the ansatz into the Lagrangian of Eq. (2) with potential (1), and applying the VA procedure, similar to that which was used for the solitons in the circular-lattice model [14], we arrive at the following evolution equation for the width of the localized state:

$$\frac{d^2 a}{dt^2} = \frac{4(1-\chi)}{a^3} + 2p\exp(-a^2)\left[-2a\exp(a^2) - \sqrt{\pi}\,\mathrm{erfi}(a) + 2\sqrt{\pi}\,a^2\mathrm{erfi}(a),\right] \quad (6)$$

where $\mathrm{erfi}(a) \equiv \mathrm{erf}(i\,a)/i$, with erf the standard error function, and

$$\chi \equiv -\alpha N\sqrt{1-\delta^2}/(4\pi), \quad (7)$$

while the amplitude is expressed in terms of the width, $A^2 = N\sqrt{1-\delta^2}/(\pi a^2)$. Recall that, as above, $\alpha = +1$ and $\alpha = -1$ correspond to the repulsive and attractive nonlinearity, respectively.

Equation (6) is tantamount to the equation of motion for a unitary-mass particle in the following external potential,

$$u(a) = \frac{2(1-\chi)}{a^2} + 2p\sqrt{\pi}\,a(t)\exp(-a^2)\mathrm{erfi}(a). \quad (8)$$

For different values of the lattice depth, *p*, the effective potential is plotted in Figs. 9(a) and 9(b), which correspond to the attractive and repulsive models, respectively. It is concluded from these results that, in the attractive model with $p=0.5$, the solitons can be localized around $r=0$ in the range of $0.45 \leq \chi < 1$ ($\chi = 1$ is the critical value, beyond which the fixed point of the potential vanishes, like in the circular lattice [14]). According to Eq. (7), this range may be translated into a respective interval of values of the norm. In the repulsive model, with $p = 6$, the solitons trapped at the center exist in the range of $0 < |\chi| \leq 1.5$. Examples of numerically found stable fundamental solitons trapped at the center are shown in Fig. 10 for both signs of the nonlinearity.

## V. CONCLUSIONS

We have studied 2D localized states in the model combining the cubic self-attraction or self-repulsion and the elliptic-lattice potential in the framework of the Gross-Pitaevskii equation. This potential, which can be used in experiments with matter-wave patterns in BEC, is different from the previously considered Mathieu lattice. The same model describes the propagation of light in bulk media with an imprinted and deformed circular grating. The existence and stability regions for fundamental and vortical annular gap solitons (with vorticity $l = 1$ and $2$), trapped in the first oval trough of the lattice potential, have been identified in the repulsive model. All the fundamental annular solitons ($l = 0$) are stable, while the vortical solitons are quasi-stable (unless their norm is too small), featuring the rocking motion in the trough. With the decrease of the norm, the shape of the annular solitons changes from almost uniform to saddle-shaped, and eventually to the DS (the double soliton, which consists of two separated in-phase peaks). The minimum value of the norm necessary for the (quasi-)stability of the vortical annular solitons with $l \neq 0$ increases with the eccentricity of the underlying lattice structure. All solitons in the repulsive model belong to the first finite bandgap, in terms of the 1D linear spectrum induced by the flat periodic lattice potential far from the center.

In the attractive model with the elliptic-lattice potential, only DS solutions exist in the first annular trough. All such solitons carrying zero vorticity are stable, while, for $l = 1$ and $2$, the DSs exhibit the quasi-stability, similar to that in the repulsive model. Stable elliptic-shaped fundamental solitons trapped in the central potential well have been found too, by means of the variational approximation and in the numerical form.

**ACKNOWLEDGMENTS**

This work was supported by Hong Kong Baptist University, the Hong Kong Research Grants Council, and the Guangdong Province branch of the Natural Science Foundation of China (Grant No. 9451063301003516).

**References**

[1] T. L. Gustavson, P. Bouyer, and M. A. Kasevich, Phys. Rev. Lett. **78**, 2046 (1997); B. P. Anderson, K. Dholakia, and E. M. Wright, Phys. Rev. A **67**, 033601 (2003).

[2] J. Fujita, S. Mitake, and F. Shimizu, Phys. Rev. Lett. **84**, 4027 (2000).

[3] J. A. Sauer, M. D. Barrett, and M. S. Chapman, Phys. Rev. Lett. **87**, 270401 (2001); S. Wu, W. Rooijakkers, P. Striehl, and M. Prentiss, Phys. Rev. A **70**, 013409 (2004); S. Gupta, K. W. Murch, K. L. Moore, T. P. Purdy, and D. M. Stamper-Kurn, Phys. Rev. Lett. **95**, 143201 (2005).

[4] E. M. Wright, J. Arlt, and K. Dholakia, Phys. Rev. A **63**, 013608 (2001).

[5] F. K. Abdullaev, B. B. Baizakov, S. A. Darmanyan, V. V. Konotop, and M. Salerno, Phys. Rev. A **64**, 043606 (2001); I. Carusotto, D. Embriaco, and G. C. LaRocca, *ibid.* **65**, 053611 (2002); B. B. Baizakov, V. V. Konotop, and M. Salerno, J. Phys. B **35**, 5105 (2002); E. A. Ostrovskaya and Y. S. Kivshar, Phys. Rev. Lett. **90**, 160407 (2003).

[6] B. Eiermann, T. Anker, M. Albiez, M. Taglieber, P. Treutlein, K. P. Marzlin, and M. K. Oberthaler, Phys. Rev. Lett. **92**, 230401 (2004); O. Morsch and M. Oberthaler, Rev. Mod. Phys. **78**, 179 (2006).

[7] B. B. Baizakov, B. A. Malomed, and M. Salerno, Europhys. Lett. **63**, 642 (2003); Phys. Rev. A **70**, 053613 (2004); J. Yang and Z. H. Musslimani, Opt. Lett. **28**, 2094 (2003); D. Mihalache, D. Mazilu, F. Lederer, Y. V. Kartashov, L. C. Crasovan, and L. Torner, Phys. Rev. E **70**, 055603(R) (2004); T. Mayteevarunyoo, B. A. Malomed, B. B. Baizakov, and M. Salerno, Physica D **238**, 1439 (2009).

[8] H. Sakaguchi and B. A. Malomed, Phys. Rev. A **75**, 013609 (2007); *ibid*. A **78**, 063606 (2008); Y. V. Kartashov, B. A. Malomed, and L. Torner, *ibid.* **75**, 061602 (R) (2007).

[9] Y. V. Kartashov, V. A. Vysloukh, and L. Torner, Phys. Rev. Lett. **93**, 093904 (2004); Y. J. He,

B. A. Malomed, and H. Z. Wang, Phys. Rev. A **76**, 053601 (2007); Y. J. He, B. A. Malomed, D. Mihalache, and H. Z. Wang, J. Phys. B: At. Mol. Opt. Phys. **41,** 055301 (2008).

[10] D. Mihalache, D. Mazilu, F. Lederer, B. A. Malomed, Y. V. Kartashov, L. C. Crasovan, and L. Torner, Phys. Rev. Lett. **95**, 023902 (2005).

[11] Y. V. Kartashov, R. Carettero-Gonzaléz, B. A. Malomed, V. A. Vysloukh, and L. Torner, Opt. Express **13**, 10703 (2005).

[12] X. Wang, Z. Chen, and P. G. Kevrekidis, Phys. Rev. Lett. **96**, 083904 (2006).

[13] L. C. Crasovan, G. Molina-Terriza, J. P. Torres, L. Torner, V. M. Perez-Garcia, and D. Mihalache, Phys. Rev. E **66**, 036612 (2002); L.-C. Crasovan, V. Vekslerchik, V. M. Pérez-García, J. P. Torres, D. Mihalache, and L. Torner, Phys. Rev. A **68**, 063609 (2003); T. Mizushima, N. Kobayashi, and K. Machida, *ibid.* **70**, 043613 (2004); M. Möttönen, S. M. M. Virtanen, T. Isoshima, and M. M. Salomaa, *ibid.* **71**, 033626 (2005); H. Pu, L. O. Baksmaty, S. Yi, and N. P. Bigelow, Phys. Rev. Lett. **94**, 190401 (2005).

[14] B. B. Baizakov, B. A. Malomed, and M. Salerno, Phys. Rev. E **74**, 066615 (2006).

[15] J. Scheuer and B. Malomed, Phys. Rev. A **75**, 063805 (2007).

[16] A. A. Zozulya, D. Z. Anderson, A. V. Mamaev, and M. Saffman, Europhys. Lett. **36**, 419 (1996).

[17] C. Rotschild, O. Cohen, O. Manela, M. Segev, and T. Carmon, Phys. Rev. Lett. **95,** 213904 (2005); Peng Zhang, Jianlin Zhao, C. Lou, X. Tan, Y. Gao, Q. Liu, D. Yang, J. Xu, and Z Chen, Opt. Express **15**, 537 (2007); S. Lopez-Aguayo and J. C. Gutiérrez-Vega, Opt. Express **15**, 18326 (2007); P. Zhang, J. Zhao, F. Xiao, C. Lou, J. Xu, and Z. Chen, Opt. Express **16**, 3865 (2008); A. Ruelas, S. Lopez-Aguayo, and J. C. Gutiérrez-Vega, Opt. Lett. **33**, 2785

(2008).

[18] A. A. Svidzinsky and A. L. Fetter, Phys. Rev. Lett. **84**, 5919 (2000); J. J. Garcia-Ripoll and V. M. Perez-Garcia, Phys. Rev. A **64**, 013602 (2001); J. J. Garcia-Ripoll, G. Molina-Terriza, V. M. Perez-Garcia, and L. Torner, Phys. Rev. Lett. **87**, 140403 (2001).

[19] F. Ye, D. Mihalache, and B. Hu, Phys. Rev. A **79**, 053852 (2009).

**Figure captions**

Fig. 1. (Color online) Elliptic lattices with (a) $\delta = 0.1$ and (b) $\delta = 0.8$.

Fig. 2. Existence regions of annular gap solitons in the first oval trough of the elliptic potential, shown in the plane of norm $N$ and ellipticity $\delta$ of lattice potential (1) with $p = 30$, in the model with the self-repulsive nonlinearity. Double solitons (DSs) exist only in region A, while elliptic annular solitons (EASs) exist in regions B, C, and D. There are no solitons at $N > N_{cr}$ for all vorticities considered, $l = 0$, 1, and 2 (if the repulsive nonlinearity is too strong, the solitons cannot be held in the single annular trough). All zero-vorticity solitons, with $l = 0$, are stable, and EASs are quasi-stable in region D for both $l = 1$ and $l = 2$, developing the rocking motion, as explained in the text. EASs are unstable in region B for $l = 1$, and in regions B and C for $l = 2$, suffering radiation losses due to the rotational motion, and eventually splitting (see further details in the text).

Fig. 3. (Color online) Examples of stable gap solitons localized in the first oval trough, in the repulsive model with $p = 30$ and $l = 0$. Strong ellipticity ($\delta = 0.8$): EASs with norm $N = 1900$ (a) and $N = 658$ (b); DS with $N = 26$ (c). Weak ellipticity ($\delta = 0.2$): EASs with norm $N = 950$ (d) and $N = 100$ (e), and DS with $N = 2$ (f).

Fig. 4. (Color online) Further results for the annular and double solitons (EASs and DSs) with zero vorticity ($l = 0$) in the model with the self-repulsion and $\delta = 0.4$. (a) The DSs and EASs exist, respectively, in regions A and B, in the plane of the lattice strength, $p$, and chemical potential, $\mu$. Examples of the stable evolution of various solitons in the lattice with strength $p = 40$ : (b) EAS with $\mu = -11.4$, (c) EAS with $\mu = -21.5$; (d) DS with $\mu = -54.4$.

Fig. 5. (Color online) Annular and double vortical gap solitons with $l = 1$ in the repulsive model with $\delta = 0.4$. (a) In the plane of $(p, \mu)$, stable DSs and EASs exist, respectively, in regions A and C, while unstable EASs are found in region B. (b) The evolution of a quasi-stable EAS with $\mu = -20.3$. (c) An example of an unstable EAS with $\mu = -23.5$. (d) A quasi-stable DS with $\mu = -55.7$. (e) Contour plots of the DS evolution corresponding to (d), which display the rocking motion of the vortical soliton, which periodically switches between the anti-clockwise and clockwise directions in the oval trough. The rocking range is between points A and B in (e). All examples of the evolution are shown here and in Fig. 6 for $p = 40$.

Fig. 6. (Color online) (Color online) Annular and double vortical solitons with $l = 2$ in the repulsive model with $\delta = 0.4$. The meaning of panel (a) is the same as in Fig. 5. (b) The evolution of a quasi-stable EAS with $\mu = -24.7$. (c) An example of an unstable EAS with $\mu = -30.5$. (d) A quasi-stable DS with $\mu = -57.4$. (e) The same as in Fig. 5(e).

Fig. 7. The maximum norm admitting the existence of DSs with $l = 0$, 1, and 2 in the attractive model, as a function of ellipticity $\delta$, for the lattice depths (strengths) $p = 10$ and 30, respectively. The upper limit, determined by the onset of the collapse, is the same for all these values of $l$.

Fig. 8. (Color online) Examples of stable double solitons in the attractive model with $N = 4.3$, $\delta = 0.4$, and $l = 0$, for different depths of the lattice: (a) $p = 10$ and (b) $p = 30$.

Fig. 9. The variational potential with $\delta = 0.1$, as given by Eq. (8), for different values of $\chi$. The strength of the elliptic lattice, $p$, is indicated in the panels.

Fig. 10. (Color online) Fundamental solitons trapped at the center of the elliptic potential with $\delta = 0.1$, in the attractive (a) and repulsive (b) models. The other parameters are $p = 0.5$, $\chi = 0.9$, and $\mu = -3.4168$ in (a), and $p = 6$, $\chi = -0.23$, and $\mu = 1.2$ in (b). Panel (c) displays the evolution of the stable soliton from (b) perturbed by random noise.

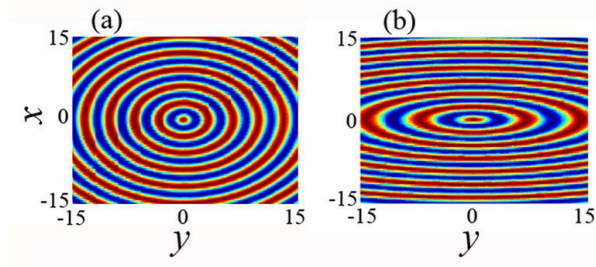

Fig. 1. (Color online) Elliptic lattices with (a) $\delta = 0.1$ and (b) $\delta = 0.8$.

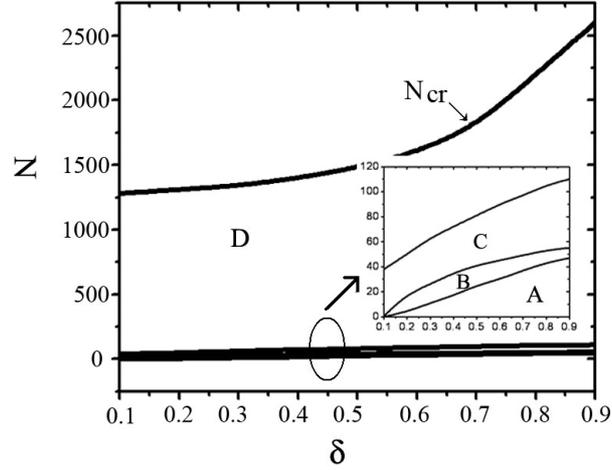

Fig. 2. Existence regions of annular gap solitons in the first oval trough of the elliptic potential, shown in the plane of norm $N$ and ellipticity $\delta$ of lattice potential (1) with $p = 30$, in the model with the self-repulsive nonlinearity. Double solitons (DSs) exist only in region A, while elliptic annular solitons (EASs) exist in regions B, C, and D. There are no solitons at $N > N_{cr}$ for all vorticities considered, $l = 0$, 1, and 2 (if the repulsive nonlinearity is too strong, the solitons cannot be held in the single annular trough). All zero-vorticity solitons, with $l = 0$, are stable, and EASs are quasi-stable in region D for both $l = 1$ and $l = 2$, developing the rocking motion, as explained in the text. EASs are unstable in region B for $l = 1$, and in regions B and C for $l = 2$, suffering radiation losses due to the rotational motion, and eventually splitting (see further details in the text).

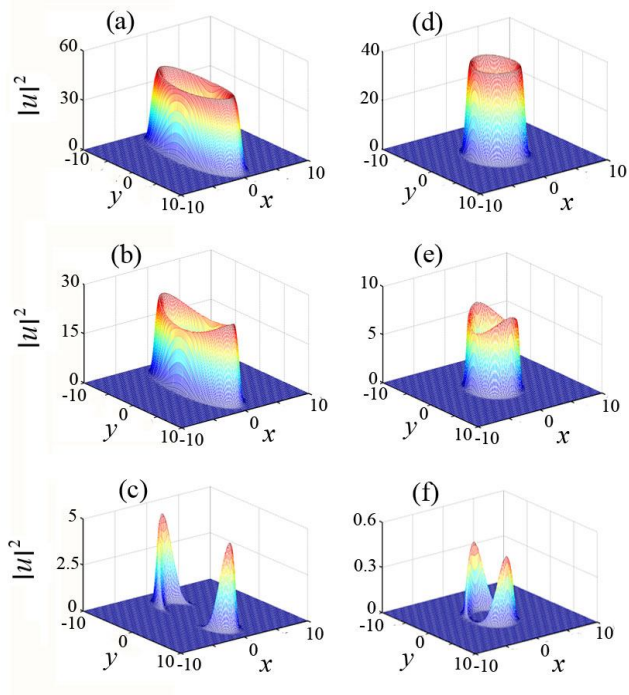

Fig. 3. (Color online) Examples of stable gap solitons localized in the first oval trough, in the repulsive model with $p = 30$ and $l = 0$. Strong ellipticity ($\delta = 0.8$): EASs with norm $N = 1900$ (a) and $N = 658$ (b); DS with $N = 26$ (c). Weak ellipticity ($\delta = 0.2$): EASs with norm $N = 950$ (d) and $N = 100$ (e), and DS with $N = 2$ (f).

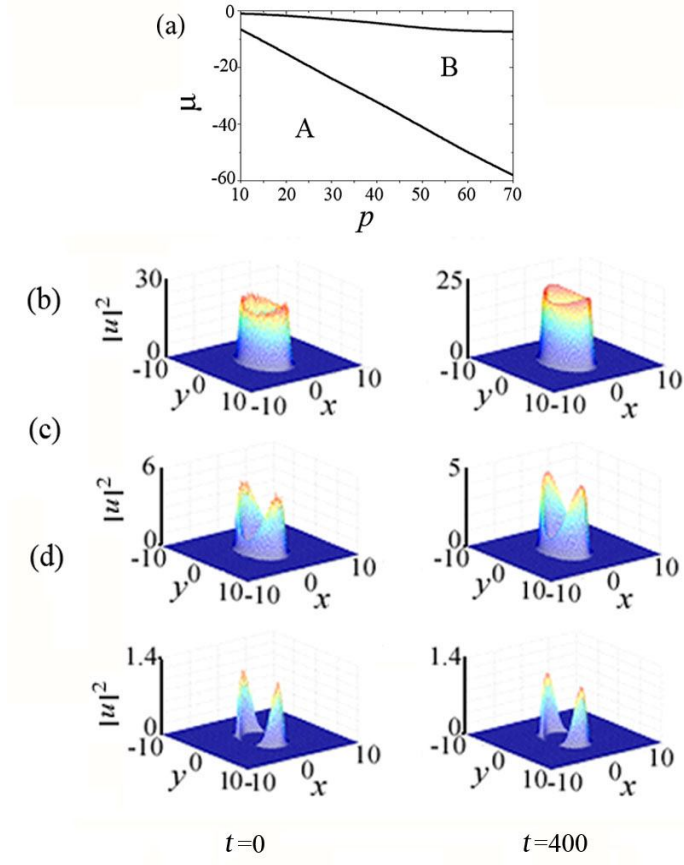

Fig. 4. (Color online) Further results for the annular and double solitons (EASs and DSs) with zero vorticity ($l = 0$) in the model with the self-repulsion and $\delta = 0.4$. (a) The DSs and EASs exist, respectively, in regions A and B, in the plane of the lattice strength, $p$, and chemical potential, $\mu$. Examples of the stable evolution of various solitons in the lattice with strength $p = 40$: (b) EAS with $\mu = -11.4$, (c) EAS with $\mu = -21.5$; (d) DS with $\mu = -54.4$.

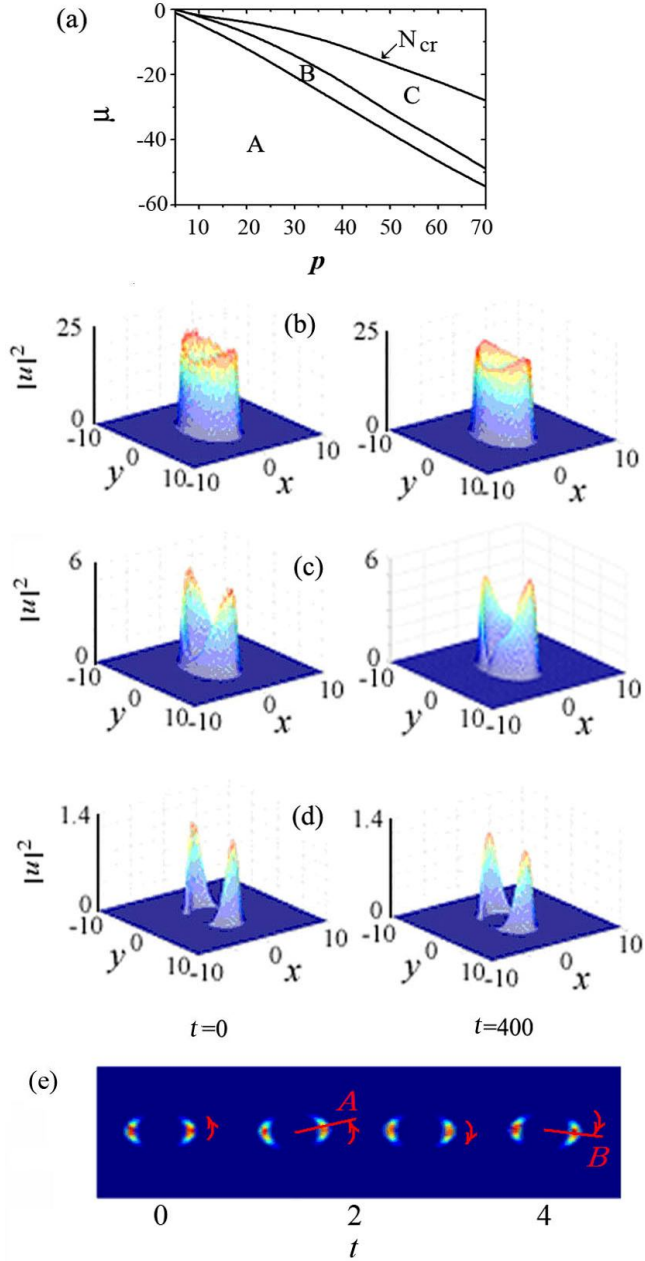

Fig. 5. (Color online) Annular and double vortical gap solitons with $l = 1$ in the repulsive model with $\delta = 0.4$. (a) In the plane of $(p, \mu)$, stable DSs and EASs exist, respectively, in regions A and C, while unstable EASs are found in region B. (b) The evolution of a quasi-stable EAS with $\mu = -20.3$. (c) An example of an unstable EAS with $\mu = -26$. (d) A quasi-stable DS with $\mu = -55.7$. (e) Contour plots of the DS evolution corresponding to (d), which display the rocking motion of the vortical soliton, which periodically switches between the anti-clockwise and clockwise directions in the oval trough. The rocking range is between points A and B in (e). All examples of the evolution are shown here and in Fig. 6 for $p = 40$.

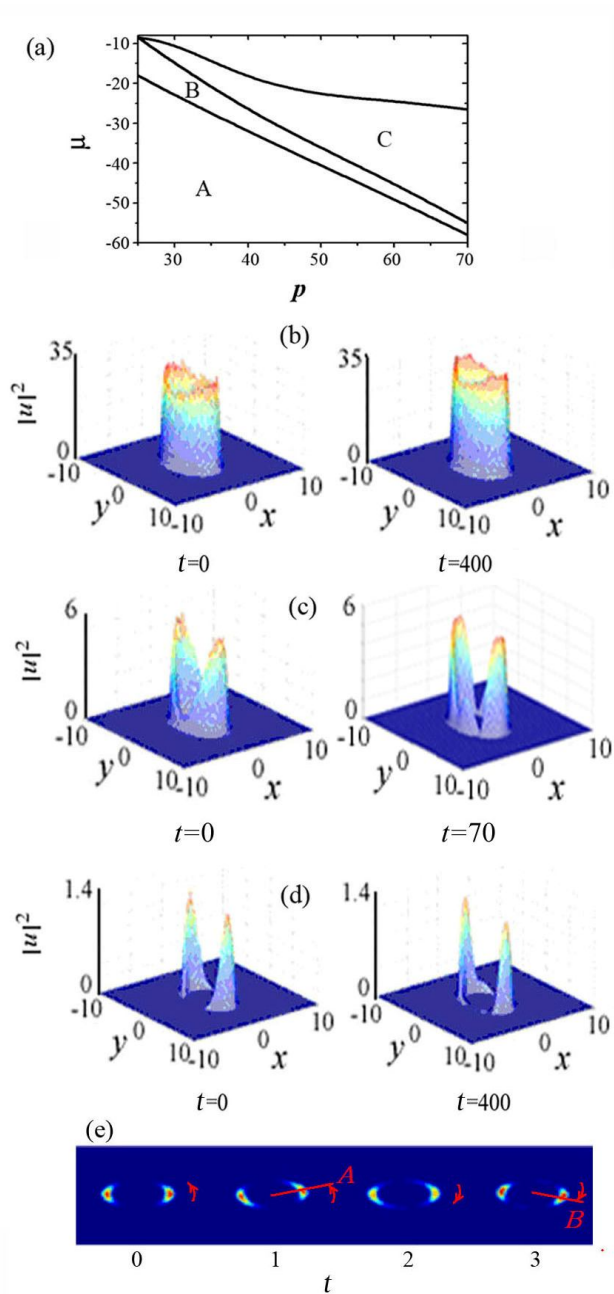

Fig. 6. (Color online) (Color online) Annular and double vortical solitons with $l = 2$ in the repulsive model with $\delta = 0.4$. The meaning of panel (a) is the same as in Fig. 5. (b) The evolution of a quasi-stable EAS with $\mu = -24.7$. (c) An example of an unstable EAS with $\mu = -30.5$. (d) A quasi-stable DS with $\mu = -57.4$. (e) The same as in Fig. 5(e).

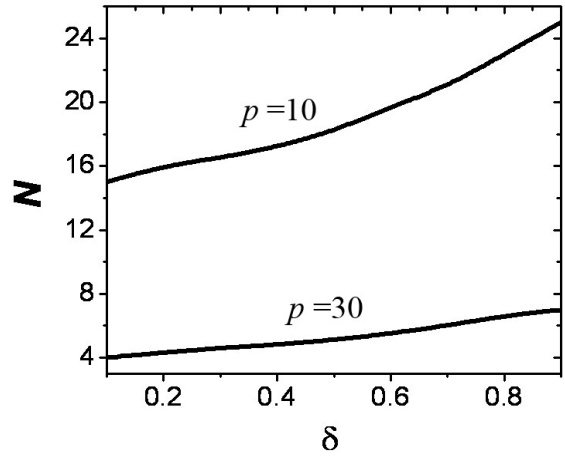

Fig. 7. The maximum norm admitting the existence of DSs with $l = 0$, 1, and 2 in the attractive model, as a function of ellipticity $\delta$, for the lattice depths (strengths) $p = 10$ and 30, respectively. The upper limit, determined by the onset of the collapse, is the same for all these values of $l$.

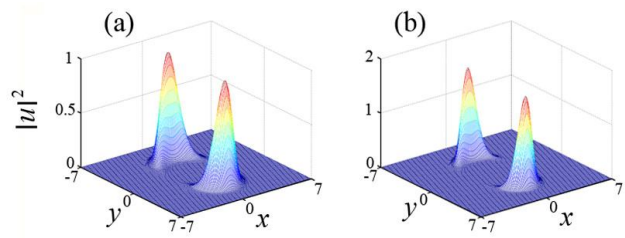

Fig. 8. (Color online) Examples of stable double solitons in the attractive model with $N = 4.3$, $\delta = 0.4$, and $l = 0$, for different depths of the lattice: (a) $p = 10$ and (b) $p = 30$.

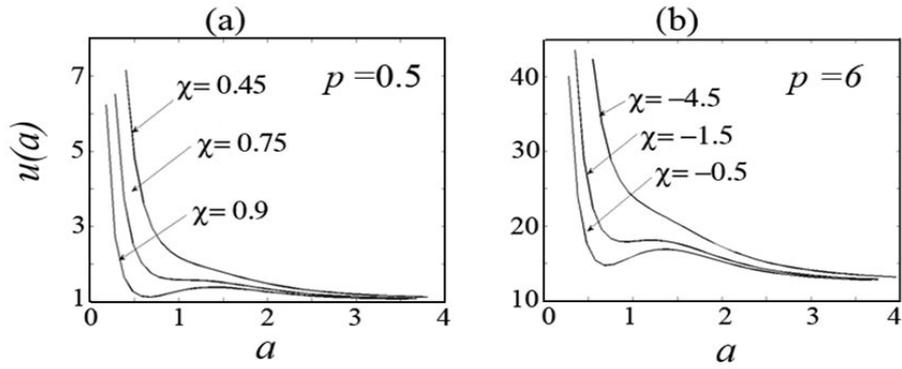

Fig. 9. The variational potential with $\delta = 0.1$, as given by Eq. (8), for different values of $\chi$. The strength of the elliptic lattice, *p*, is indicated in the panels.

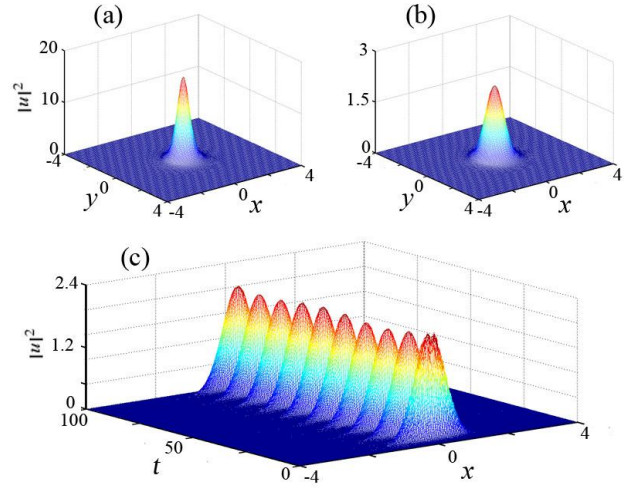

Fig. 10. (Color online) Fundamental solitons trapped at the center of the elliptic potential with $\delta = 0.1$, in the attractive (a) and repulsive (b) models. The other parameters are $p = 0.5$, $\chi = 0.9$, and $\mu = -3.4168$ in (a), and $p = 6$, $\chi = -0.23$, and $\mu = 1.2$ in (b). Panel (c) displays the evolution of the stable soliton from (b) perturbed by random noise.